\documentclass[11pt,titlepage]{article}

\usepackage{setspace}

\usepackage{subfigure}

\usepackage{amsmath}

\usepackage{graphicx}

\usepackage[round,numbers,sort&compress]{natbib}
\title{Folding a protein with equal probability of being helix or hairpin}

\author{Chun-Yu Lin\\
    Department of Physics, \\
    National Tsing-Hua University, Hsinchu 300, Taiwan
    \and Nan-Yow Chen \\
    National Center for High-Performance Computing, Hsinchu 300, Taiwan
    \and Chung~Yu Mou\thanks{
           Corresponding author.  Address:
           Department of Physics,
       National Tsing Hua University,
       101 Section 2 Kuang Fu Road,
       Hsinchu, Taiwan~30013, R.O.C.,
       Tel.:~(886)3-574-2537, Fax:~(886)3-572-3052} \\
    Department of Physics, \\
    National Tsing Hua University, Hsinchu 300, Taiwan \\
    Institute of Physics, Academia Sinica, Nankang, Taiwan\\
   Physics Division, National Center for Theoretical Sciences, Hsinchu 300, Taiwan}

\date{}

\begin{document}

\maketitle

\abstract{We explore the possibility for the native state of a protein being inherently
a multi-conformation state in an ab initio coarse-grained model.
Based on the Wang-Landau algorithm, the complete free energy landscape for the
designed sequence 2D4X: INYWLAHAKAGYIVHWTA is constructed.
It is shown that 2DX4 possesses two nearly degenerate native states: one has a helix structure, while the other
 has a hairpin structure and their energy difference is less
than 2\% of that of local minimums.  Two degenerate native states
are stabilized by an energy barrier of the order 10kcal/mol. Furthermore, the hydrogen-bond
and dipole-dipole interactions are found to be two major competing interactions
in transforming one conformation into the other. Our results indicate that  degenerate native
states are stabilized by subtle balance between different interactions in proteins; furthermore,
for small proteins, degeneracy only happens for proteins of sizes being around 18 amino acids or 40 amino acids.
These results provide important clues to the study of native structures of proteins.

\emph{Key words:} Wang-Landau algorithm 1; $\alpha$-$\beta$
transition 2; coarse grained model 3}

\clearpage

\section*{Introduction}

Solving the protein folding problem has tremendous implications.
Among possible applications, the solution to the problem makes it
possible to design drugs theoretically, which would result in the
greatest impact to the biological science. Nonetheless, despite much
effort being devoted during the past, the problem continues to be
one of the most basic unsolved problems. To solve the folding
problem completely, it is generally believed that to be able to
predict the protein structure for a given sequence of amino acids is
an important step. This belief originates from the classical
Anfinsen's work \citep{Anfinsen} and is often summarized by stating
that there is a unique native configuration for a given sequence of
amino acids. Over the past decades, this point of view, however, has
been challenged by experimental evidences. It is now known that
proteins can be driven to different folded states by changing pH
value, ionic strength \citep{Cerpa1996,Zhang1997}, temperature
\citep{Murayama2004}, or solvent polarity \citep{Montserret2000}.
These facts indicate that there may exit nearby competing states to
the native state of globular proteins in vivo. Therefore, given
appropriate conditions, the native state of a given sequence of
amino acids can be changed. In particular, it implies that the
structure of a given segment of amino acids may depend on  the
context it resides in.

Indeed, accumulating evidences have indicated that  the secondary structure can
be context dependent. Bovine $\beta$-lactoglobulin protein is a predominantly
$\beta$-sheet protein but it has been observed to go through a remarkable
$\alpha$ to $\beta$ transition during the folding process
\citep{Hamada1996,Kuwata1998}.   Kabsch and Sander also found a pentapeptide
sequence which could adopt an $\alpha$-helical or a $\beta$-sheet
conformation in different proteins. Cohen and colleagues
\citep{Cohen1993} extended this work to hexapeptides. Minor and Kim
\citep{Minor1996} have conducted an experiment showing that an 11 amino acid
sequence can be transformed into an $\alpha$-helical or a $\beta$-sheet
in protein G. Such 'chameleon' sequences have their
cooperative local interactions competing against long range
interactions of sequence environment. The fragmental propensity of
secondary structures are found to be overwhelmed by larger
structures.

To elucidate the mechanism that causes the conformation change, a \emph{de
novo} protein has recently been designed \citep{Araki2007}. The
modified sequence INYWLAHAKAGYIVHWTA posited in Protein Data Bank
\citep{Bernstein1978} (PDB ID 2DX3 and 2DX4, and we shall term it simply as 2DX4
hereafter) from residues 101-111 of human $\alpha$-lactalbumin
was identified to has equal population of $\alpha$-helical or
$\beta$-sheet in an aqueous solution. Although it is well recognized
that protein solutions are in equilibrium with intermediate
peptides, the dual native states are rarely reported in the
literature. Furthermore, it is shown that the conformational transformation of 2DX4
is not induced by any environmental
conditions or binding motifs. These facts make 2DX4 a valuable target to study.
In particular, folding 2DX4 would be a crucial test for any viable approach for
solving the protein folding problem.

On the theoretical side, although all-atom simulation is the most
comprehensive approach for understanding the folding processes;
nonetheless, the requirement of computational resources tends to be
realistically unaffordable \citep{Duan}. Itoh, Tamura and Okamoto
\citep{Itoh2010} have combined all-atom molecular dynamics
simulation with multi-canonical multioverlap algorithm to simulate
2DX4. From the limited phase space obtained, they investigated
possible pathways for the $\alpha$ to $\beta$ transition. In
particular, three local minima in free energy are identified.
However, only partial $\alpha$ helix or $\beta$ hairpin are found in
the structures associated with these local minima. The mechanism
that is responsible for the possibility of two native states of 2DX4
thus remains unclear. On the other hand, there have been much effort
in developing coarse grained models to predict protein structures
\citep{Thirumalai}. In these models, effects of water molecules are
implicitly included in effective interacting potentials between
amino acids. The required computational resources is much reduced
and it enables the prediction of protein structures feasible.
Indeed,  progress have recently been made in predicting structures
of wild-type proteins of sizes from 12 to 56 amino acids by using
realistic and unbiased potentials between amino acids
\citep{Chen2006}.  To further check the validity of coarse grained
models, folding proteins such as 2DX4 would be an ideal test.

In this work, based on the ab initio coarse-grained model
constructed in Ref. \citep{Chen2006}, we constructed the complete
free energy landscape for 2DX4. It is shown that in agreement with
the experimental observation, there are only two local minima with
structures being $\alpha$-helix and $\beta$-hairpin respectively.
Moreover, within the accuracy of the coarse-grained model, it is
found that while two local minima are degenerate in the case of
2DX4, the $\beta$-hairpin  is higher in energy for the DP3 protein
which results from the mutation of one amino acid of 2DX4 and was
reported to have zero population of hairpin structure
\citep{Araki2007}. In addition, the pathways between the helix and
hairpin configurations are simulated by Monte Carlo (MC) algorithm
in high temperatures. By analyzing detailed free energy profile, we
find that the hydrogen-bond and dipole-dipole interactions are two
major competing mechanism in transforming one conformation into the
other. Our results indicate that  generally, degenerate native
states are stabilized by subtle balance between different
interactions in proteins. Furthermore, for small proteins,
degenerate native states only happen for proteins of sizes being
around 18 amino acids or 40 amino acids. These results provide
important clues to the study of native structures of proteins.

\section*{Theory and Methods}

\subsection*{Ab Initio Coarse-grained Potentials }

\begin{figure}
   \begin{center}
      \includegraphics*[width=1.0\textwidth]{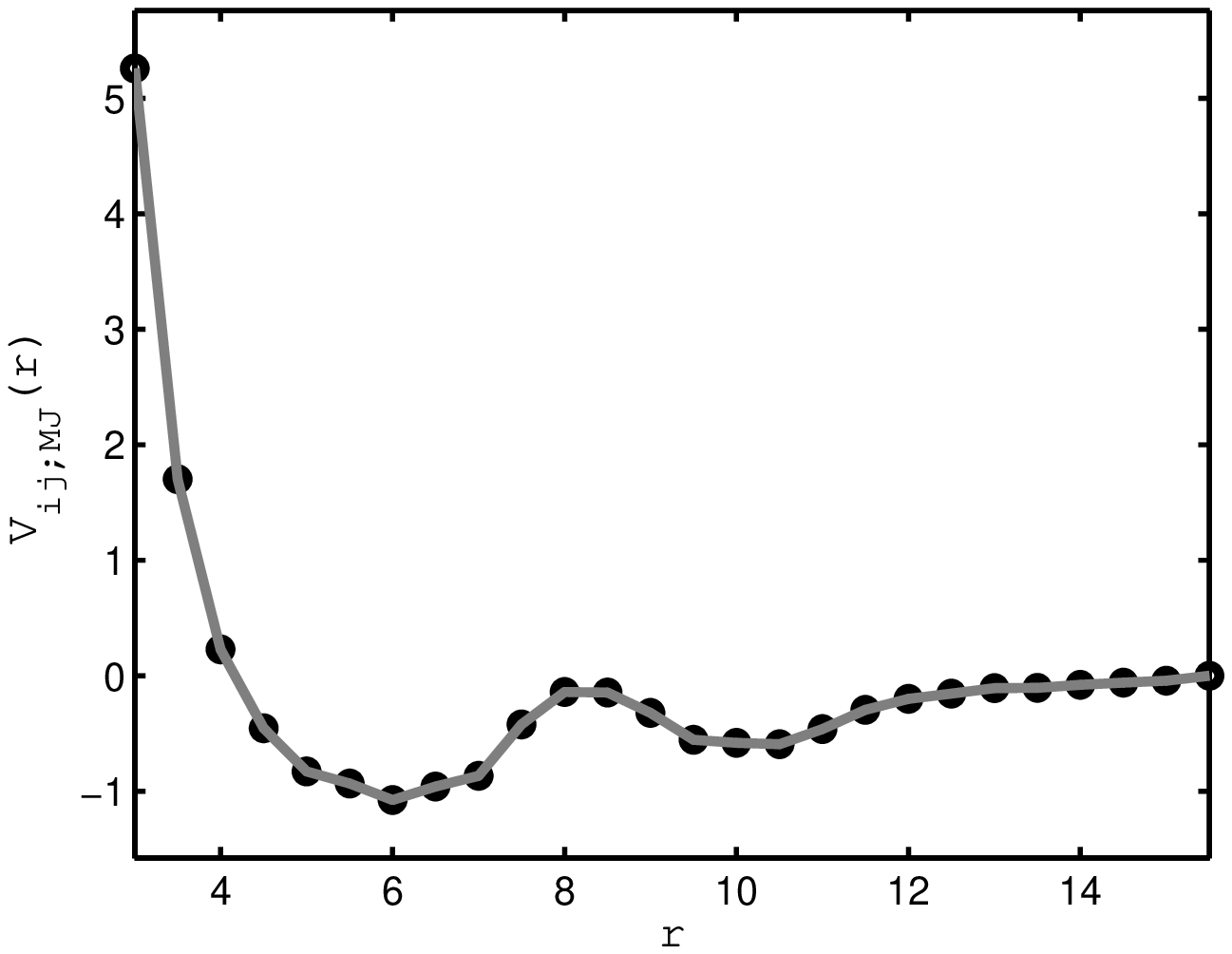}
      \caption{A typical effective potential, $V_{ij;MJ}(r)$. Here the potential is  between Valine and Leucine and the solid line is the continuous curve interpolated between data obtained by statistical analysis of PDB. One sees that even though there are structures in proteins, $V_{ij;MJ}(r)$ shows liquid-like behavior and exhibits similar oscillations in consistent with the desolvation model.}
      \label{fig:fig1}
   \end{center}
\end{figure}

We shall first recapture essentials of the coarse-grained model
constructed in Ref. \citep{Chen2006}. In this model, residues are
coarse-grained as spheres centered at $\rm{C}^\beta$ atoms but
complete structures are kept in backbones.  Bond angles and bond
lengths are fixed between these atoms to increase folding
efficiency, the only variables are dihedral angles $\phi$ and $\psi$
on the $\rm{C}^\alpha$ atom hinging two amide planes. On the other
hand, water molecules are not included explicitly but their effects
are incorporated in effective potentials among side-chains and
backbones. In these representations and
 with all energies being in unit of kcal/mol,
 the total energy can be written as
\begin{equation} \label{eqn:Eterms}
   E_{total} = E_{Steric} + E_{DD} + E_{HB} + E_{MJ} + E_{NP} + E_{SA}.
\end{equation}
Here each energy term is a weighted potential energy with $E_i =
\varepsilon_i V_i$ , where $\varepsilon_i$ is the weighting factor
to be determined later and $V_i$ is the corresponding potential
energy. Among these energy terms, $E_{Steric}$ is to enforce the
structural constraints such as hard-core potentials to avoid
unphysical contacts, while $E_{SA}$ is solvent accessible surface
energy in proportional to the area of each side-chain that is
exposed to water and is primarily responsible for stabilizing the
tertiary structure. The remaining terms are three ingredients for
the formation of the secondary structures with $E_{HB}$ being the
hydrogen bonding between any non-neighboring $NH$ and $CO$ pair,
$E_{DD}$ being the summation of screened dipole-dipole interaction
at large distance (global dipole interaction, $E_{DG}$) and local dipole-dipole interaction
between dipoles on the backbones, and $ E_{MJ} + E_{NP}$ accounting
for the interactions due to hydrophobicity or the charge state of
the amino acids. All the potentials were based on realistic
parameters obtained from experimental data except for $ E_{MJ} +
E_{NP}$, which was based simple generalizations of Miyazawa-Jernigan
matrix \citep{Miyazawa1985,Miyazawa1996} by using 12-6 Lennard-Jones
potential modified by effects due to sizes of water molecules
\citep{Chen2006}. In order to include realistic effects due to
hydrophobicity or the charge state of the amino acids, we shall
construct the corresponding potentials by statistical methods so
that $ E_{MJ}$ generalizes Miyazawa-Jernigan Matrix
\citep{Miyazawa1985,Miyazawa1996} to finite large distances between
amino acids, while $E_{NP}$ generalizes the $V_{LocalHP}$ in Ref.
\citep{Chen2006} and is the statistical energy that characterizes
the propensity (to $\alpha$ or $\beta$) of amino acids in nearest
neighbors. With these potentials, the weighting factors
$\epsilon_i$'s are calibrated based on a few proteins of known
structures \citep{Chen2006}. Their values are $\varepsilon_{DG}$
=0.21,  $\varepsilon_{DN}$ = 2.0, $\varepsilon_{HB}$ = 4.8,
$\varepsilon_{SA}$ = 1.35,  and  $\varepsilon_{MJ}$ = 0.85; while
for helix and sheet propensity energies, we get
$\varepsilon_{NP}^{\alpha}$ = 6.4 and  $\varepsilon_{NP}^{\beta}$ =
16,

To extend the Miyazawa-Jernigan Matrix to finite distances, we perform
extended statistical analysis by first writing
\begin{equation} \label{eqn:EMJ1}
   E_{MJ} = \varepsilon_{MJ}\sum_{i, j}V_{ij;MJ}(r)(1 - SA_i)(1 - SA_j),
\end{equation}
where $SA_i$ and $SA_j$ are the solvent accessibilities for $i$th
and $j$th residue respectively. The quantity, $V_{ij;MJ}(r)$, is the
statistical potential between the $i$th and $j$th residues obtained
by counting number $n_{ij}$, of the corresponding $i$-type and
$j$-type residues separating by $r$, that appears in the PDB.
Fundamentally, $V_{ij;MJ}(r)$ is the generalization of the pair
distribution function \citep{Chaikin} and its relation
 to $n_{ij}(r)$ is given by the Boltzmann's statistics
\begin{equation} \label{eqn:EMJ_rdf}
   \exp(-V_{ij;MJ}(r)) = A(r_k)\frac{\sum\limits_{p}n_{ij;p}(r_k)}{\sum\limits_{p,r_k}\frac{\displaystyle (n_{ir;p}(r_k)+n_{i0;p}(r_k))(n_{jr;p}(r_k)+n_{j0;p}(r_k))}{\displaystyle
   (n_{rr;p}(r_k)+n_{r0;p}(r_k))}},
\end{equation}
where $A(r_k)$ is a normalization factor to be determined later,
numbers with the index $p$ denote the corresponding statistical
values that belong to one specific protein $p$, $0$ represents the solvent group, and
$r_k$ is the radius of the kth spherical shell centered at
$i$-type residue. Note that different amino acids have different
occurrence frequency in real proteins and this is normalized by
the denominator in Eq.\ref{eqn:EMJ_rdf}. Furthermore, homology of sequence bias was eliminated by
the sequence alignment method in combination with the weighting matrix used by
Miyazawa and Jernigan \citep{Miyazawa1996}.  Here 2$n_{ij} (r_k)$ for i
$\neq$ j and $n_{ii} (r_k)$ are the counts when the $i$-type residue is at the origin and the $j$-type residue
is in the kth distance $r_k$, while
$n_{ir}$ is the total count of the $i$the residue
\begin{equation} \label{eqn:nir}
   n_{ir;p}(r_k)=\sum\limits_{j}n_{ij;p}(r_k).
\end{equation}
$n_{i0}$ counts events taking place between the $i$-type residue and
solvent group 0
\begin{equation} \label{eqn:ni0}
   n_{i0;p}(r_k)=\frac{1}{2}q_i(r_k)n_{i;p}(r_k)-n_{ir;p}(r_k),
\end{equation}
 where $q_i$ is the coordinate number of the $i$-type residue in the $k$th spherical shell and $n_i$
is the total number of the $i$-type
residues in protein $p$. $n_{rr}$ and $n_{r0}$ are
summations of $n_{ir}$ and $n_{i0}$ over $i$-type residue
respectively
\begin{eqnarray}
   n_{rr;p}(r_k)=\sum\limits_{i}n_{ir;p}(r_k), \label{eqn:nrr1} \\
   n_{0r;p}(r_k)=\sum\limits_{i}n_{i0;p}(r_k). \label{eqn:nrr2}
\end{eqnarray}
Finally, the
normalization factor A(r) is defined by
\begin{equation} \label{eqn:Anorm}
   A(r_k)= \frac{{\rm total\hspace{2pt}number\hspace{2pt}of\hspace{2pt}shells}}{\frac{4}{3}\pi[(r_k + \frac{\Delta r}{2})^3 - (r_k - \frac{\Delta r}{2})^3]},
\end{equation}
where $\Delta r$ is the width of each spherical shell. The effective
potential as a continuous function of $r$, $V_{ij;MJ}(r)$, is then
interpolated from $V_{ij;MJ}(r_k)$. As a demonstration, in
Fig.~\ref{fig:fig1} ,  we show a typical effective potential
obtained by the above statistical analysis. We see that similar to
the pair distribution function for liquid molecules \citep{Chaikin},
$V_{ij;MJ}(r)$ exhibits similar oscillations in consistent with the
desolvation model \citep{desolvation}. Furthermore, even though
there are structures in proteins, there is no indication of any
ordering in $V_{ij;MJ}(r)$.

\begin{figure}
   \begin{center}
      \vbox{
          \subfigure[]{\includegraphics[width=1.0\textwidth]{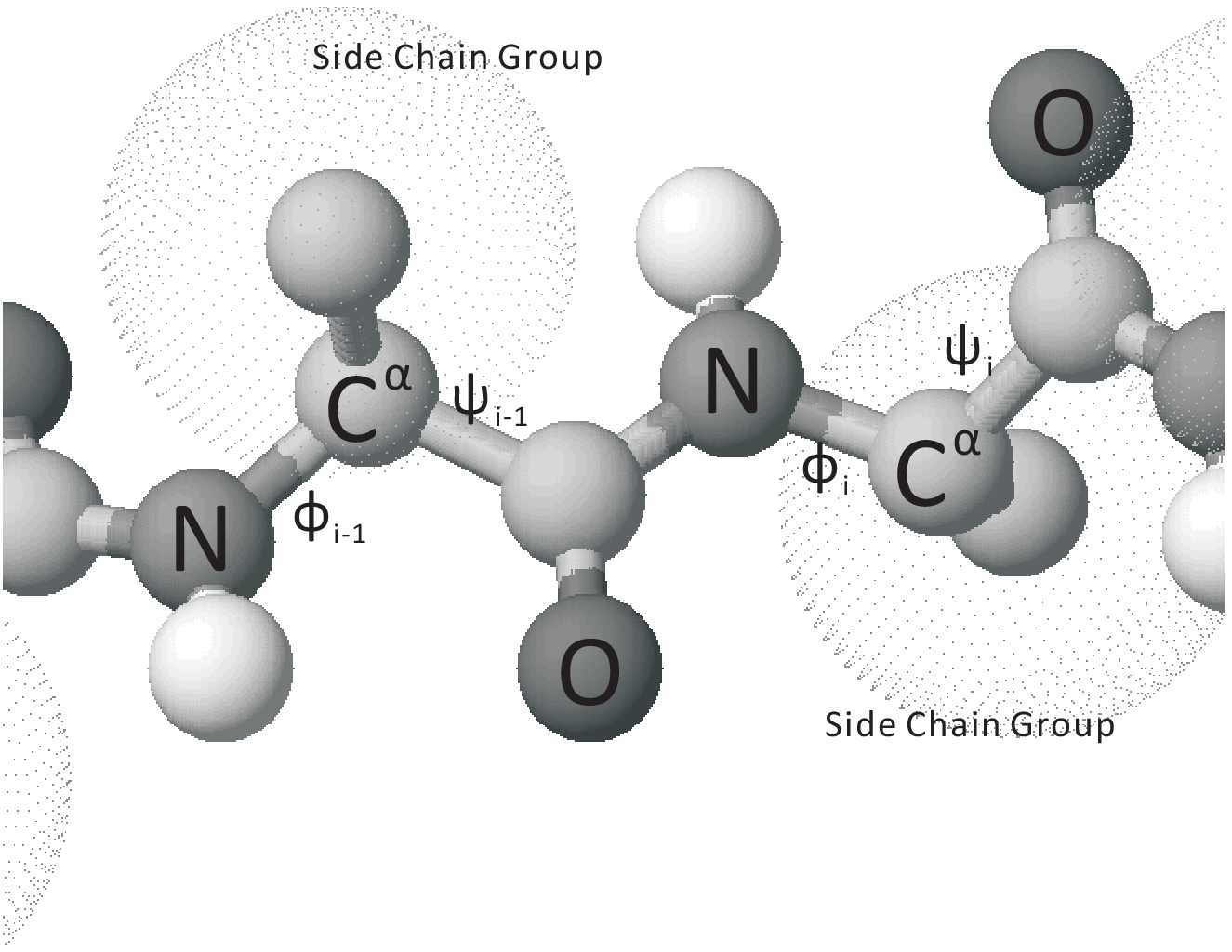}
         \label{fig:fig2a}
           }
           \subfigure[]{\includegraphics[width=1.0\textwidth]{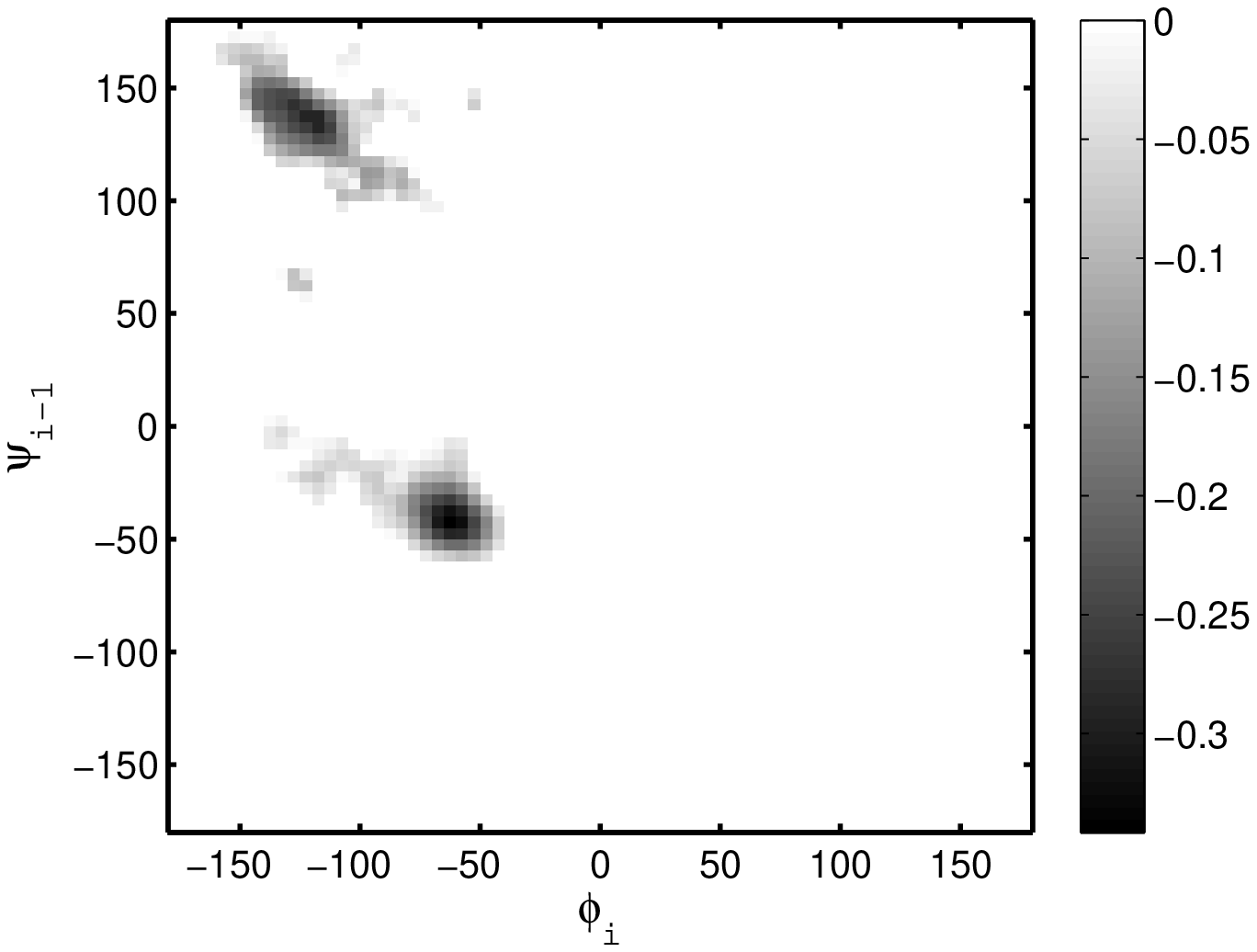}
            \label{fig:fig2b}
          }
        }
      \caption{(a) Dihedral angles that characterize effective potentials for nearest neighboring residues. (b) A typical effective potential, $V^{\alpha}_{i-1,i}+ V^{\beta}_{i-1,i}$, between nearest neighboring amino acids. Here the interaction is between the Aspartic acid and Tyrosin. Similar to the Ramachandran plot, the potential is significant only in the regions with $\alpha$ or $\beta$ structures.}
      \label{fig:fig2}
   \end{center}
\end{figure}

The effective $V_{ij;MJ}(r)$ is only valid for large enough distances.
For residues in nearest neighbors, due to the steric constraints,
the pair distribution function starts to deviate from the desolvation model.
To extend $E_{MJ}$ to characterize interactions of residues in nearest neighbors,  $E_{NP}$ is
introduced to account for the statistical energy between nearest neighboring residues.
The interactions among nearest neighboring residues are
best characterized by dihedral angles $\phi$ and $\psi$ of the
corresponding  amide planes. Since $V_{ij;MJ}(r)$ does not cover distances
of three successive residues, $E_{NP}$ needs to characterize three successive
residues in the protein, labeled by $i-1$, $i$, and $i+1$. Using the corresponding dihedral angles
shown in Fig.~\ref{fig:fig2a} ,
$E_{NP}$ can be written as
\begin{equation} \label{eqn:ENP1}
   E_{NP}= \sum\limits_{i}\sum\limits_{k=\alpha,
   \beta}\varepsilon_{NP}^k[V_{lm}^k(\psi_{i-1}, \phi_i) + V_{mn}^k(\psi_i, \phi_{i+1})]V_{m}(\phi_i,
   \psi_i),
\end{equation}
where $l$, $m$ and $n$ are indices for type of residues, $V_m$
is a one-body potential that depends on $\psi_i$ and $\phi_i$ of the
amide planes connecting to the $m$-type residue, and $V_{lm}$ (also $V_{mn}$) is a
two-body energy that depends dihedral angles of $l$-type and $m$-type residues
in nearest neighbors. According to the Ramachandran plot, it is known that
$\phi$ and $\psi$ are statistically concentrated at
particular regions, which are either in the $\alpha$-helix configuration
or $\beta$-sheet configuration. To ensure the
relative magnitudes of $\alpha$-helix and $\beta$-sheet part are
not biased by the database, different weighting factors with $k=\alpha$ and $\beta$
are introduced in Eq.~\ref{eqn:ENP1}.

The one-body angular potential $V_m$ is obtained by first analyzing the
bare potential $v_m$ defined by
\begin{equation} \label{eqn:E1body}
   \exp(-v_m(\phi, \psi))=\frac{n_m(\phi, \psi)}{\int\int n_m(\phi, \psi)d\phi d\psi},
\end{equation}
where $n_m$ is the number density taken over the whole PDB for type $m$ residues with dihedral angles being $(\phi, \psi)$. To account for the preference or non-preference
of $\alpha$ or $\beta$ structures, we set
$V_m(\phi_i, \psi_i) = \theta (\Lambda - v_m(\phi_i,
\psi_i) )$ with $\theta$ being the step function and $\Lambda$ being a negative threshold 
energy level so that $V_m$ is either $1$ or $0$.

The bare two-body potential is constructed by
\begin{equation} \label{eqn:E2bodyStatistic}
\exp(-v_{lm}^k(\psi_{i-1}, \phi_i))=\frac{n_{lm}(\psi_{i-1}, \phi_i)\int\int n_{rr}(\psi_{i-1}, \phi_i)d\psi_{i-1} d\phi_i}
   {\int\int n_{lr}(\psi_{i-1}, \phi_i)d\psi_{i-1} d\phi_i \int\int n_{mr}(\psi_{i-1}, \phi_i)d\psi_{i-1} d\phi_i},
\end{equation}
where $n_{lr}$, $n_{mr}$ and $n_{rr}$ are defined in
same way as those in Eq.~\ref{eqn:nir} and Eq.~\ref{eqn:nrr1} except that they are specialized to
the dihedral angle $(\psi_{i-1}, \phi_i)$.
$V_{lm}^k(\psi_{i-1}, \phi_i)$ is then defined by rescaling $v_{lm}$ with respect to
the average value of $v_{lm}$
\begin{equation} \label{eqn:E2body}
   V_{lm}^k(\psi_{i-1}, \phi_i)= (A_{lm}^k - A_{ave}^k) v_{lm}^k(\psi_{i-1}, \phi_i) /
   A_{lm}^k
\end{equation}
where $A_{lm}$ is the minimum of $v_{lm}$ over all possible $(\psi_{i-1}, \phi_i)$
and $A_{ave}$ is the average value of $A_{lm}$ over all possible pairs of amino acids. A typical $V_{lm}$ is shown in Fig.~\ref{fig:fig2b} . It is clear that $V_{i-1,i}(\psi_{i-1}, \phi_i)$ does not vanish only in particular regions, in which local structures of proteins are either $\alpha$ helices or $\beta$ sheets.

\subsection*{Wang-Landau Monte Carlo algorithm}

Given the ab initio coarse-grained potential obtained, one can
determine the free energy landscape by using the  Wang-Landau
algorithm \citep{Wang2001a}. The density of states is estimated by
random sampling on energy space via the transition probability
\begin{equation} \label{eqn:Prob}
   P(E_1\rightarrow E_2)=\min(\frac{g(E_1)}{g(E_2)}, 1),
\end{equation}
where $g(E)$ is the density function of energy E. Although this
algorithm was first demonstrated on Ising model of
spin array, it is portable to molecular systems with continuous
energy value \citep{Rathore2003, Ojeda2009a}. Specific
implementations adapted in our work are the following steps:
\\
(1) Define a density function g(E, X) and histogram H(E, X)
with X's being any variables
other than energy.  Set initial values: g(E,X) = 1 and H(E, X) = 0 for all E and X.
\\
(2) Generate an initial conformation randomly and
calculate its energy $E_1$.
\\
(3) Generate a new conformation by making a small change (e.g. the
dihedral angles). Calculate the new energy $E_2$ and the transition to the new conformation
is determined by the transition probability
$P(E_1, X_1 \rightarrow E_2, X_2) = \min[g(E_1,X_1)/g(E_2, X_2), 1]$.
\\
(4) If  the system stays in the original $E_1$ state,  $g(E_1, X_1)$ is replaced by
$g(E_1, X_1) \times f$ and $H(E, X)$ is accumulated through $H(E_1,
X_1) + 1$. Otherwise, one sets $g(E_2, X_2) = g(E_2, X_2) \times f$ and $H(E_2, X_2)
= H(E_2, X_2) + 1$. The factor f is initially set to $e^1$.
\\
(5) After each MC step, check if less than 2 \% of sites in $H$ are
smaller than flat threshold, which is defined to be 10 \% of averaged
$H(E, X)$. If this is satisfied, the histogram is flat and one then sets $f =
\sqrt{f}, H(E, X) = 0$ and goes to step (2).  When $f < \exp(10^{-3.6})$ is satisfied,
one exists the procedure.
\\

All the above steps are identical to Wang-Landau's scheme except for the flat
histogram criteria in step (5), which is modified to accommodate enormous states
involved for proteins so that sampling can be done in finite computation time.
Once the density of states is constructed, the free energy landscape can
be calculated as
\begin{equation} \label{eqn:Prob}
   F(E,X)=E-k_BT\log[g(E,X)],
\end{equation}
where $k_B$ is the Boltzmann constant and $T$ is the absolute temperature. The
variable space X is not restricted to be one dimension and has to be chosen
to exhibit the landscape.

\section*{Results}

\subsection*{Propensity Analysis and Monte Carlo Simulation}
\begin{figure}
   \begin{center}
      \includegraphics*[width=1.0\textwidth]{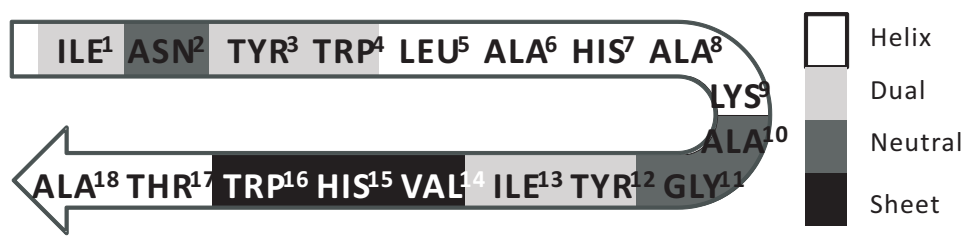}
      \caption{Nearest neighbor propensity of 2DX4 obtained by statistical analysis of PDB.
Here the dual propensity implies the residue pair can adopt  either $\alpha$ or $\beta$
structure. By contrast, the neutral propensity implies that the residue pair is free
to rotate in dihedral angles and it is often that a turn
region of anti-parallel $\beta$-sheet is developed.}
      \label{fig:fig3}
   \end{center}
\end{figure}

To investigate the energy landscape of  2DX4, we first analyze its propensity.
Past studies \citep{Betancourt2004, Matsuo1994} have indicated that
each amino acid has its propensity of secondary structure.
By using the constructed statistical potential $V_{lm}$ (see Theory and Methods),
we summarize the nearest neighbor propensity of 2DX4 in Fig.~\ref{fig:fig3}.
Here amino acids in nearest neighbors
are classified according to the tendency of corresponding amino acids
being in $\alpha$-helix, $\beta$-sheet, dual or neutral.
The dual propensity implies the residue pair can adopt  either $\alpha$ or $\beta$
structure. By contrast, the neutral propensity implies that the residue pair is free
to rotate in dihedral angles and it is often that a turn
region of anti-parallel $\beta$-sheet is developed. From the propensity analysis,
it is clear that even though there is no absolute global tendency for 2DX4
being $\alpha$ helix or $\beta$ sheet, by including residues with neutral and dual propensities,
there are more residues in favor of $\alpha$ helix.  Nonetheless,
the high $\beta$-sheet propensity near the C-terminal, containing amino acids
V, H, and W, indicates the possibility of switching 2DX4
between helix and hairpin structures. Since each of these three amino acids
has larger side chain radius than the averaged radius of others, it is more difficult for the
segment to curl into part of the helix structure. As a result, the strand formed by residues 14-18 regularly dangles in solvent and deposits a nucleation seed to transform from $\alpha$-helix
to $\beta$-sheet.

\begin{figure}
   \begin{center}
      \includegraphics*[width=1.0\textwidth]{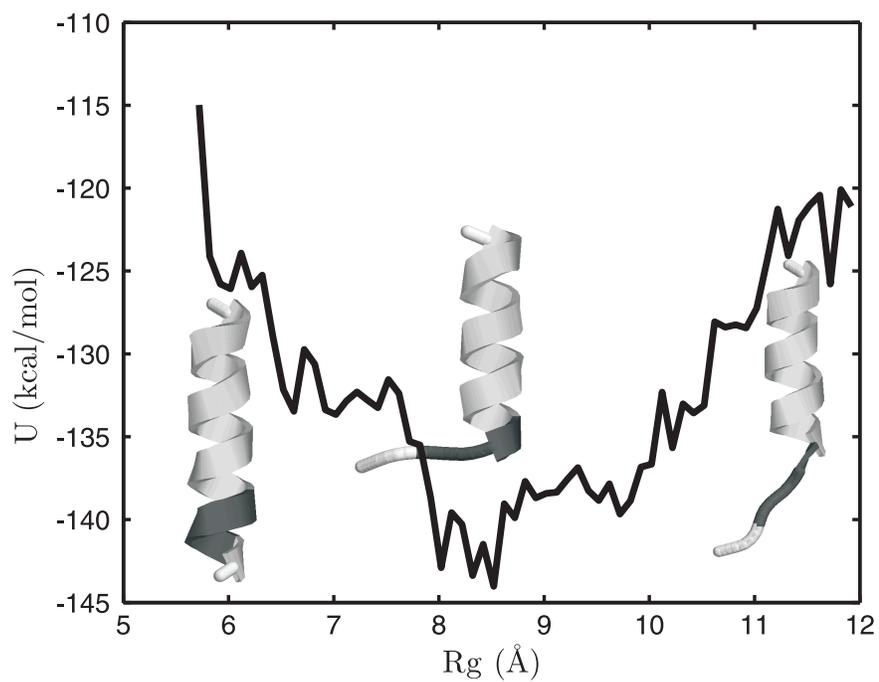}
      \caption{Internal energy U versus the radius of gyration Rg for $\alpha$-like structures.
Due to the dangling motion of the strand VHW near C-terminal, the complete helix is not the lowest energy state. The protein snapshots are drawn by RasMol \citep{Sayle1995}.}
      \label{fig:fig4}
   \end{center}
\end{figure}

In order to investigate the stability of $\alpha$ helix due to
residue 14-18, a MC simulation of 2DX4 by starting from an all helix
conformation is conducted. Since the expanding of the strand affect
the size of 2DX4, we record the radius of gyration ($Rg$) for
structure resembling the $\alpha$-helix. Larger $Rg$ represents
structures with extended strands, while smaller $Rg$ represents
structures which are closer to the standard $\alpha$-helix. Since
each $Rg$ interval may contains several helix structures with
different energy values, the internal energy $U$, defined by the
Boltzmann statistics $U = \sum_{E}E\exp(-\beta E)$, is evaluated as
a function of Rg. In Fig.~\ref{fig:fig4}, we show the plot of $U$
versus $Rg$. It is seen that the lowest energy state is not a
complete $\alpha$ helix.  In general,  hydrogen bonds and long range
dipole energy favor helix structures \citep{Chen2006}.  In the case
of 2DX4, nearest neighbor interactions $V_{NP}$ compete with these
helix-favored energies and result in the lowest total energy state
with partial helix and partial strand structure. The native
structure found in our MC simulation is identical to results
obtained by the experiment \citep{Araki2007} and other simulations
\cite{Itoh2010}, indicating the credibility of the coarse-grained
potentials  described in Eq. \ref{eqn:Eterms}.

To clarify the final fate of $\alpha$ helix, we perform full MC
simulations by starting from the initial state of a straight line
with all dihedral angles $\phi$ and $\psi$ being equal to 180
degree.  Indeed, two configurations of lowest energies are found and
correspond to $\alpha$ helix and $\beta$ hairpin with RMSD (root
mean square deviation of positions) being 3.74 ${\AA}$ and 4.40
${\AA}$ respectively. The simulations take $4 \times 10^8$ MC steps
and ended on either helix or hairpin states. Furthermore, starting
from an $\alpha$ helix at 400 K (RT = 0.8 kcal/mol),  the $\alpha$
helix is transformed into a $\beta$ sheet and vice versa. All of the
transitions occurred successfully in our MC simulations. However,
the helix to hairpin transition takes twice to ten times of more MC
steps than that for the transition from hairpin to the helix. A
hairpin to helix transition finished approximately in $5 \times
10^7$ MC steps, where the reverse process took $10^8$ MC steps or
longer. The obtained asymmetry of transition rates are consistent
with the literature report \citep{Munoz1997}, where progress of
$\beta$-hairpin formation is thirty times slower than the rate of
$\alpha$-helix.

\subsection*{Free Energy Landscape}

In order to make sure if helix and hairpin structures found in MC
simulations are the only two minimum, we calculate the free energy
by employing the Wang-Landau algorithm. In addition to the energy,
we characterize the energy landscape by using the contact ratio $Q$
as another coordinate. Here $Q$ is defined by using the conformation
of the minimum state with helical like structure as the reference
state so that $Q$ is the ratio of contact number of the state to
that of the minimum state. The free energy $F$ is thus  a function
of energy $E$ from  -150 kcal/mol to 0 kcal/mol and contact ratio Q,
ranging from 0 to 1.  In the calculation, to insure that all regions
can be accessed,  a trial run with $4 \times 10^8$ MC steps is first
performed to identify regions with scarce probability. In the latter
runs, free energy density in these regions will be computed
separately.

\begin{figure}
   \begin{center}
        \vbox{
            \subfigure[]{\includegraphics[width=1.0\textwidth]{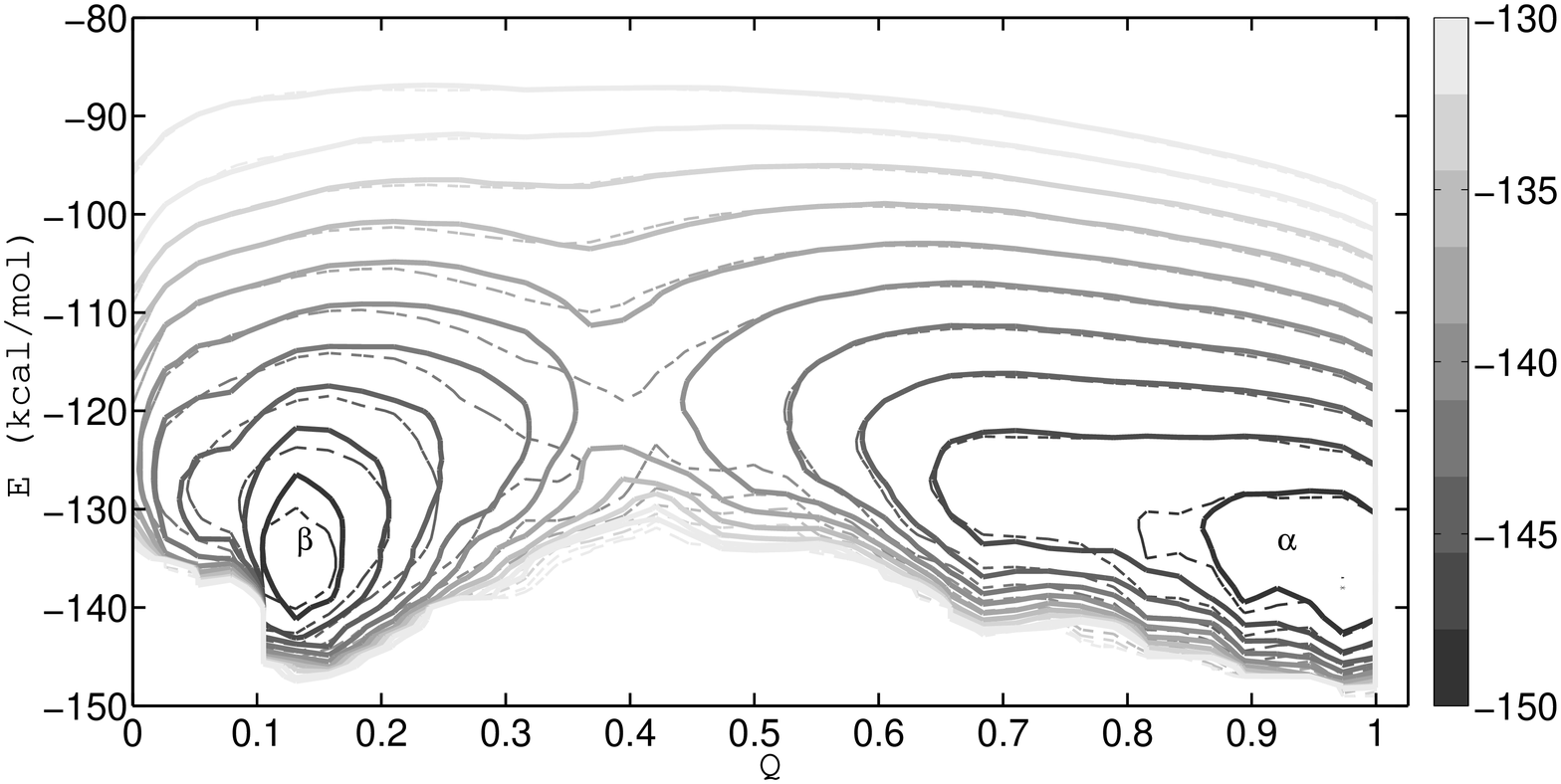}
            \label{fig:fig5a}
            }
            \subfigure[]{\includegraphics[width=1.0\textwidth]{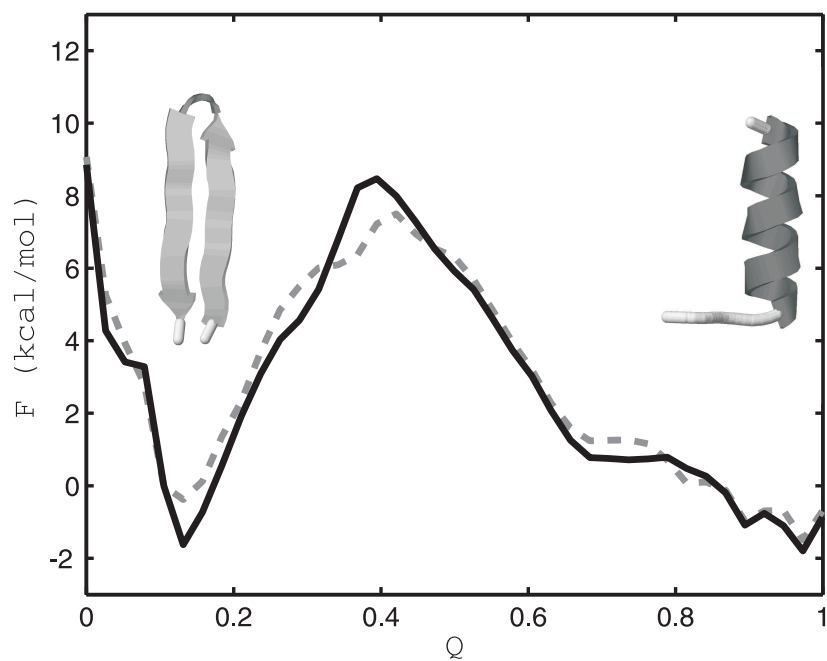}
            \label{fig:fig5b}
            }
        }
        \caption{(a) Free energy contour $F(E, Q)$ for 2DX4 (solid lines) and DP3 (dash lines)
at the experimental temperature, 283 K. Here $Q$ is the contact ratio and  DP3 is mutated 2DX4 through Y12S mutation.  Two minima with helix-like and hairpin structures labeled by $\alpha$ and $\beta$ are exhibited for both cases; however, for DP3, helix region gets expanded, while hairpin region gets shrunk, indicating that helix structure is more stable for DP3. (b) Free energy curves $F(Q)$ for 2DX4 and DP3. The helix structure becomes the most stable structure for DP3, in consistent with experiments.}
        \label{fig:fig5}
   \end{center}
\end{figure}

Figure~\ref{fig:fig5a} shows the resulting complete free energy
landscape for 2DX4. It demonstrates that the free energy has only
two minimum at helix and hairpin states. The difference of free
energies for helix and hairpin structures is less than 0.17 kcal/mol
at room temperature, which clearly demonstrates that 2DX4 is a
two-state protein with two stable native states. In
Fig.~\ref{fig:fig5b}, the one-dimensional free energy curves $F(Q)$
are deduced from the density of states $g(E, Q)$ via the formula
$\exp(-F(Q)/KT) = \sum_{E}g(E, Q)\exp(-E/KT)$. A free-energy barrier
around 10 kcal/mol exists between helix and hairpin structures,
Since the energy barrier is much larger than typical energy
fluctations $k_B T$,  it stabilizes both helix and hairpin. The free
energy landscape also depends on temperature. At temperature $k_B T
= 0.8$, about 400 K, the minimum at helix side expands from $Q = 1$
to $Q = 0.65$ with residues 1-10 being kept in helix conformation.
In other words, half of the peptide on N-terminal is thermally
stable in helix, and residues 11-14 are free to denature at high
temperatures.

As a comparison, we examine energy landscapes of  mutated 2DX4
through Y12S mutation, which are labeled as DP3 and DP5 in the
previous experiment \citep{Araki2007}.  It is reported that DP3 has
zero population of hairpin formation in the sense that even though
there is minor intra-strand signal, there is no inter-strand signal
for the hairpin structure. It is therefore important to examine
native states of DP3 in the current model.  Figure~\ref{fig:fig5a}
reveals that for DP3, helix region gets expanded, while hairpin
region gets shrunk. This indicates that helix structure is more
stable for DP3. Indeed, Fig.~\ref{fig:fig5b} shows that the free
energy of the helix state is less than that of the hairpin state by
1.1 kcal/mol at room temperature. In addition, we find that this
energy difference is sensitive to temperature and becomes 1.4
kcal/mol at 100 K. In contrast, for DP5, the free energy of the
helix state is found to be fixed at 100-298K, suggesting that
helical structure is thermally more stable in DP5 than in DP3, in
agreement with experimental observation \citep{Araki2007}.  Note
that it is presumed \citep{Araki2007} that absence of $\pi$-$\pi$
interaction of Tyr12-His7 near the turn region is the cause for the
absent of hairpin in DP3. However, close examination based on the
propensity  indicates that mutation of Y12S in DP3 intensify the
$\beta$ sheet propensity of the second strand. Thus the lack of
hairpin population in DP3 is due to inter-strand interactions not
intra-strand propensity. These results are consistent with
experimental results that DP3 has only intra-strand signal. The
rigidity of second strand and absent of one $\pi$-$\pi$ bond are
thus responsible for unstable helix as well as zero population of
hairpin in DP3.

\begin{figure}
   \begin{center}
      \includegraphics*[width=1.0\textwidth]{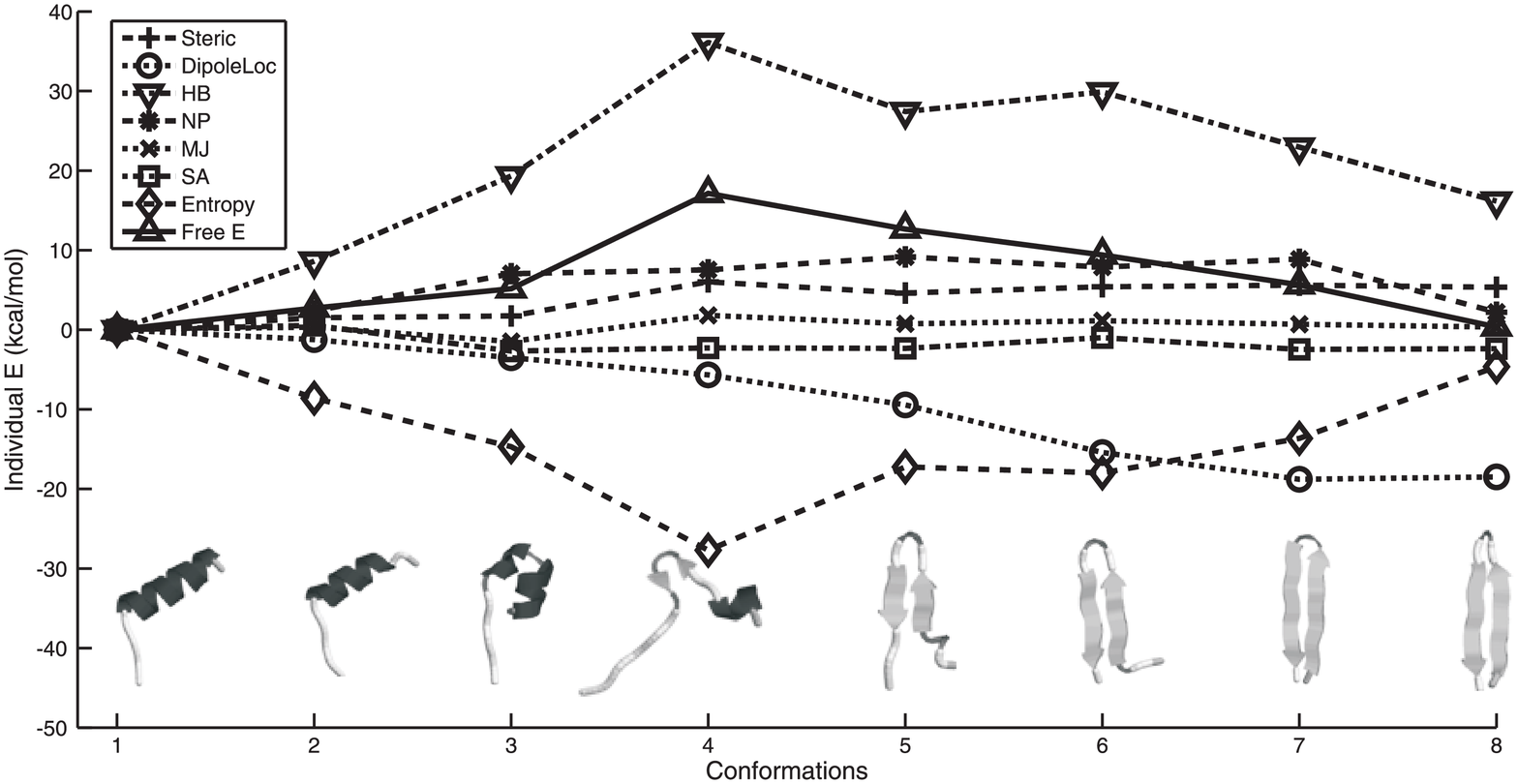}
      \caption{Comparison of different energy contribution during the transition between the helix and
the hairpin structures. Here the entropy is defined by $k_B \log [g(E,Q)]$. Large compensation
between hydrogen bond energy (HB) and local dipole energy indicates that compromising of HB and local dipole energy are the mechanism for the occurrence of degenerate native states.}
      \label{fig:fig6}
   \end{center}
\end{figure}

The mechanism for the existence of degenerate native states can be
explored by analyzing changes of different energy terms when 2DX4
changes between the helix and the hairpin structures.
Figure~\ref{fig:fig6} shows changes of different energies on the
path between the helix and the hairpin structures. We find that the
degeneracy is due to a large compensation between hydrogen bond
energy (HB) and local dipole energy. Physically, it is known that
the helix structure has more hydrogen bonds \citep{Chen2006} and
hence one looses energy in hydrogen bonds by going from the helix
structure to the hairpin structure. On the other hand, $\beta$
sheets contain large anti-parallel dipoles on nearest neighboring
amide planes, which lowers down the local dipole interaction energy.
Differences of other energy terms in 2DX4 are around 2-3 kcal/mol.
Therefore, our results show that the compensation of these two
energy leads to the degeneracy of the helix and hairpin structures.


\section*{Discussion and Conclusion}

\begin{figure}
   \begin{center}
      \includegraphics*[width=1.0\textwidth]{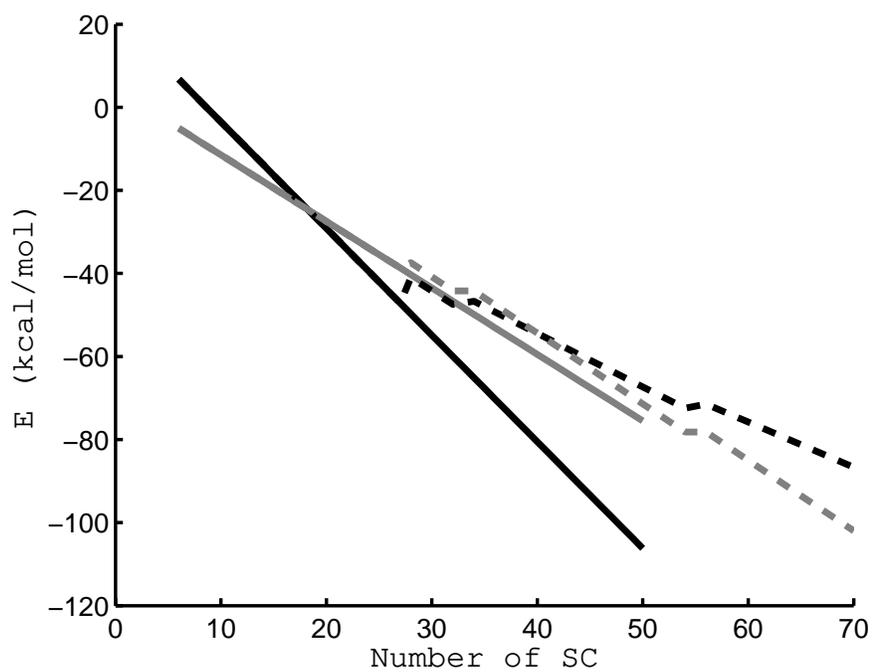}
      \caption{Optimized difference of hydrogen bond energy
and local dipole energy for $\alpha$ helix and $\beta$ sheet versus
number of side chains.  Here solid lines are differences for $\beta$ sheet being a simple hairpin.
Dash lines are optimized difference with respect to number of $\beta$ strands.}
      \label{fig:fig7}
   \end{center}
\end{figure}

In summary, the possibility for the existence of degenerate native states provides new
insight into the folding mechanism of proteins.
 Our results show that the possibility is realized in the designed 2DX4, which
possesses two nearly degenerate native state:  one has a helix structure, while the other
 has a hairpin structure.  Furthermore, the
existence of degenerate native states is driven by large
compensation between hydrogen-bond energy and local dipole energy.
The mechanism suggests that 2DX4 may not be the only protein with
degenerate native states. To examine other possibility for proteins
with degenerate native states, we examine the difference of hydrogen
bond energy and local dipole energy for $\alpha$ helix and $\beta$
sheet versus number of side chains.   The energy difference is
optimized with respect to number of $\beta$ strands.
Figure~\ref{fig:fig7} shows the computed optimized difference of
hydrogen bond energy and local dipole energy for $\alpha$ helix and
$\beta$ sheet versus number of side chains. It is seen that in
addition to 2DX4 with 18 amino acids, balance of hydrogen bond
energy and local dipole energy also happens when number of side
chains is around 40. It indicates that by suitable choice of amino
acids with balanced interactions in proteins, degeneracy can happen
for proteins of sizes being around 18 amino acids or 40 amino acids.
These results will be important clues for further construction of
proteins with degenerate native states.

\section*{Acknowledgments}
We thank Prof. Chia-Ching Chang for helpful discussions. This work was supported by National Science Council of Taiwan.


\newpage
\section*{Figures}
\clearpage
\clearpage
\clearpage

\clearpage

\clearpage
\clearpage
\clearpage

\end{document}